\documentclass[aps,prl,twocolumn,superscriptaddress,nofootinbib]{revtex4-1}

\usepackage{graphicx}
\usepackage{dcolumn}
\usepackage{bm}
\usepackage{amssymb}
\usepackage{amsmath}
\usepackage{float}
\usepackage{epsfig}
\usepackage{amsmath}
\usepackage{color}
\usepackage{txfonts}
\usepackage{microtype}
\usepackage{ulem}
\usepackage{pdfpages}

\makeatletter
\newcommand*{\rom}[1]{\expandafter\@slowromancap\romannumeral #1@}
\AtBeginDocument{\let\LS@rot\@undefined}
\makeatother

\begin{document}

\title{Semi-classical limitations for photon emission in strong external fields}

\author{E. Raicher}
\email{erez.raicher@mpi-hd.mpg.de}
\affiliation{Max-Planck-Institut f\"{u}r Kernphysik, Saupfercheckweg 1,  69117 Heidelberg, Germany }
\author{S. Eliezer}
\affiliation{Department of Applied Physics, Soreq Nuclear Research Center, Yavne 81800, Israel }
\affiliation{Nuclear Fusion Institute, Polytechnic University of Madrid, Madrid, Spain }
\author{C.H. Keitel}
\affiliation{Max-Planck-Institut f\"{u}r Kernphysik, Saupfercheckweg 1,  69117 Heidelberg, Germany }
\author{K.Z. Hatsagortsyan}
\email{karen.hatsagortsyan@mpi-hd.mpg.de }
\affiliation{Max-Planck-Institut f\"{u}r Kernphysik, Saupfercheckweg 1,  69117 Heidelberg, Germany }

\date{\today}

\begin{abstract}

The semi-classical heuristic emission formula of Baier-Katkov
%V.N. Baier, V.M. Katkov
[Sov. Phys. JETP \textbf{26}, 854 (1968)] is well-known to describe radiation of an ultrarelativistic electron in strong external fields employing the electron's classical trajectory. To find the limitations of the Baier-Katkov approach, we investigate electron radiation in a strong rotating electric field quantum mechanically using the Wentzel-Kramers-Brillouin approximation.
Except for an ultrarelativistic velocity, it is shown that an additional condition is required in order to recover the widely used semi-classical result.
A violation of this condition leads to two consequences. First, it gives rise to qualitative discrepancy in harmonic spectra between the two approaches. Second, the quantum harmonic spectra are determined not only by the classical trajectory but also by the dispersion relation of the effective photons of the external field. 
\end{abstract}

%\pacs{12.20.Ds, 52.38.-r}
\maketitle

In recent years the achievable
intensities of both optical \cite{gemini, ELI, XCELS} and x-ray \cite{LCLS, XFEL} lasers are rapidly rising and
a plethora of unexplored physical phenomena are expected to come within reach \cite{Tajima, mendonca2, DiPiazza, schwinger, macchi, esarey, poder, Cole, wistisen}. The fundamental theory describing these phenomena is strong-field QED. The strong field regime is characterized by a large  nonlinearity parameter $\xi\gg 1$ \cite{ritus}, with
$\xi \equiv ea/m = 7.5 \sqrt{ I_L/(10^{20}W/cm^2) }/\omega \left[(eV)) \right]$, where
$-e$ and $m$ are the electron charge and mass, respectively, $a$
is the amplitude of the laser vector potential $A_{\mu}$ and
$I_L$, $\omega$ are the laser intensity and frequency, respectively.
Relativistic units $\hbar = c = 1$ are used throughout.

The Furry picture is commonly employed in strong field QED, when the strong electromagnetic field is considered as a classical
field and is included in the free part of the Lagrangian \cite{Furry}. As a consequence, the free particles in the standard QED perturbation theory are replaced with
particles experiencing the external
field.
The rates of the scattering processes of electrons, positrons and photons in the presence of a laser field calculated in this framework \cite{ritus, nikishov, landau, QED1, QED2, QED3, QED4, QED5, QED6, QED7, QED8, QED9, QED10} depend upon $\xi$, as well as on the quantum parameter
$\chi \equiv \frac{e}{m^3} \sqrt{-(F^{\mu \nu} P_{\nu})^2}$,
where $P_{\nu}$ is the
kinetic momentum and $F_{\mu \nu} = {\partial}_\mu A_{\nu} - {\partial}_\nu A_{\mu}$ the
field
tensor.

Especially successful is this formalism for field configurations which admit exact analytical solution for the
electron wave function: plane wave (PW) \cite{Volkov}, static electric field \cite{Nikishov_E}, static magnetic field \cite{klepikov,sokolov}, and PW combined with a static magnetic field \cite{redmond}.
A real experiment, however, may involve
complex
field configurations, especially in plasma environment. A possible way to overcome this obstacle is to rely on the semi-classical (SC) method, introduced by Baier and Katkov \cite{katkov1, katkov2}. It allows one to calculate the emission quantum mechanical rate, including photon recoil effects, in general field configuration, given that the particle motion is quasiclassical and ultrarelativisitic.
This method does not require the particle wave function, but only its classical trajectory in the given field configuration. In the ultrastrong field limit $\xi \gg 1$,
a more simple Constant Crossed Field (CCF) approach is applicable
\cite{nikishov, elkina, DP_CCF, ilderton_CCF}.
Recently, an approximation treating
an ultrarelativistic electron interacting with a tightly focused laser beam has been proposed \cite{DPsol1} and used for calculation the rates of QED processes in such fields \cite{DPsol2, DPsol3, DPsol4}.
Given the increasingly extreme and complex scenarios being explored with further rising laser intensities and frequencies, there is urgent need to investigate the limitation of the SC approximation.

A relatively simple but not exactly solvable field configuration is the rotating electric field (REF), whose associated wave function was excessively studied
\cite{my1, my3, Varro, Heinzl_sol, Noga, becker}. Its physical significance stems from the fact that it describes a particle in the anti-node of a standing laser wave as well as a particle in a plasma wave (in a frame that moves with the wave group velocity).
Recently, quantum calculations of the emitted radiation in this configuration
have been carried out \cite{my4, Mack}, using
different approximations for the wave function.

In the present letter the quantum radiation emitted by a particle in a REF is calculated using the Wentzel-Kramers-Brillouin (WKB) method, which is more accurate than the approaches employed so far for this scenario. The full
quantum calculation of the radiation spectra is used as a benchmark for the SC approach. We demonstrate that a further condition is required except for an ultrarelativisitc motion: the $\gamma$-factor of the particle should be much
larger than the non-linearity parameter $\xi$. As a test case, a particle moving in a circle in REF is studied. As the latter takes place when the initial momentum of the electron is vanishing, the condition $\gamma\gg \xi$ is violated, and the WKB and SC approaches yield different emission spectra. According to both models,
the spectrum takes the form of discrete harmonics, but with qualitatively different harmonic structure (although, with the same harmonic-averaged spectrum).
Moreover, the WKB model predicts a cutoff in the number of absorbed photons, while the SC model shows
a semi-continuum in the high energy tail of the spectrum. As the circle-like electron's trajectory is also possible for a particle counterpropagating a PW laser with a certain initial momentum, we compare our emission spectrum to the latter case as well. For the PW (at $\xi\gg 1)$, a continuous spectrum, coinciding with the CCF result, is predicted in contrast to the discrete spectrum in REF. The observed discrepancy indicates that the quantum emission spectrum cannot be fully determined by the electron's classical trajectory.

We calculate the radiation emitted by an electron in a REF using the WKB approximation for the wave function.
For the sake of simplicity, the spin effect is neglected and thus the electron wave function is described by the Klein-Gordon equation, which is justified as long as $\chi \lesssim 1$ \cite{ritus}.
The vector potential of REF is given by
$A^{\mu} = a^{\mu}_1 \cos (k \cdot x) + a^{\mu}_2 \sin (k \cdot x)$,
where $a^{\mu}_1 = a(0,1,0,0)$, $a^{\mu}_2 = a(0,0,1,0)$ are the polarization vectors,
$k = (\omega,0,0,0)$ is the wave
vector in the laboratory frame and $\omega$ is the field frequency.
 $k^2  \neq 0$ in contrast to the PW case.
In the following the leading order WKB approximation is employed \cite{brezin, mocken}.
Namely the wave function reads $\Phi = \frac{1}{\sqrt{2\mathcal{H}(t)}}e^{iS}$ where
$S(t)=\int_{-\infty}^t{\mathcal{H}}dt-\textbf{p} \cdot \textbf{x}$ is the classical action, a bold letter designates a 3-vector and $\mathcal{H} = \sqrt{m^2 + \textbf{P}^2}$ is the Hamiltonian. The
kinetic momentum is related to the initial one by $\textbf{P} = \textbf{p}-e \textbf{A}$.
The photon emission amplitude in the Furry picture reads \cite{Supp}:
\begin{equation}
iT \propto
  \mathcal{T}_{fi}
 %{-\infty}^{\infty}
 \delta^3(\textbf{p}_{c}), \quad \mathcal{T}_{fi}
 %\textbf{p}_T), \quad \mathcal{T}_{fi}
 %{t_i}^{t_f}
   \equiv e \int_{t_i}^{t_f} dt
\left[ \frac{\epsilon \cdot P}{\sqrt{2\omega' \mathcal{H} \mathcal{H}'}} \right]
e^{i \psi}
\label{eq:av trans_amp12}
\end{equation}
where the proportion factor appears in \cite{Supp}, $\textbf{k}'$ and $\textbf{p}'$  are the momenta of the outgoing photon and electron, respectively, and the following definition was used $\textbf{p}_{c} \equiv \textbf{p} -  \textbf{k}' - \textbf{p}'$.
The exponent argument
is $\psi =  \omega' t +\int_{\infty}^t {dt'} \mathcal{H}'(t') - \int_{-\infty}^t {dt'} \mathcal{H}(t')$.
Due to momentum conservation $\textbf{p}'=\textbf{p} - \textbf{k}'$,
the time dependent momentum of the outgoing electron may be written as $\textbf{P}'(t)=\textbf{P}(t) - \textbf{k}'$.
It allows us to write $\mathcal{H}'(t) $ in terms of $p,k'$, namely
$ \mathcal{H}'(t) =  \sqrt{\left[\mathcal{H}(t) - \omega' \right]^2 - 2 P(t) \cdot k'} $.
The key approximation required in order to recover SC is:
\begin{equation}
\Xi_1 \equiv \frac{k' \cdot P(t)}{\left( \mathcal{H}(t)- \omega' \right)^2}  \ll 1.
\label{eq:av Xi1}
\end{equation}
As a result, one may Taylor expand $\mathcal{H}'(t) $.
Substituting it in the expression for $\psi$, the first and second terms cancel out and one obtains exactly the phase of the time-integrand in the
Baier-Katkov expression \cite{katkov1, katkov2}
\begin{equation}
\psi \approx \widetilde{\psi} \equiv \left( \frac{\mathcal{H}}{\mathcal{H} - \omega'} \right) \left[ k' \cdot x(t) \right],
\label{eq:av psi_SM}
\end{equation}
where the 4-momentum and energy are related by the 4-velocity $v_{\mu}$,
 $P_{\mu}=\mathcal{H} v_{\mu}$. The tilde symbol designates the SC approximation.
In order to carry out the time integration over $\mathcal{H}$, required to obtain (\ref{eq:av psi_SM}), a further assumption was made. Namely, the factor $\mathcal{H}/(\mathcal{H} - \omega')$ was regarded as a constant. In addition, in order to recover the SC prefactor of the transition amplitude for a scalar particle, the $\mathcal{H}'$ appearing in the denominator of (\ref{eq:av trans_amp12}) should be equal to $\mathcal{H}-\omega'$.
These two requirements imply that $\mathcal{H}$ should be approximately constant.
Since generally speaking, the Hamiltonian is oscillating in time, we formulate this condition as
\begin{equation}
\Xi_2 \equiv \frac{\Delta \mathcal{H}}{\mathcal{H}_{av}} \ll 1
\label{eq:av Xi2}
\end{equation}
where $\Delta \mathcal{H}$ is the deviation of $\mathcal{H}$ from the cycle average value $\mathcal{H}_{av}$.

Having obtained the mathematical requirements for the equivalence between the two approaches, let us discuss the physical conditions for which they are satisfied.
First, notice that in the classical limit ($\omega' \ll \mathcal{H}$) Eqs. (\ref{eq:av Xi1}) and (\ref{eq:av Xi2}) are both satisfied and the classical emission formula is recovered. From now on the quantum case is considered, namely $\omega'$ is smaller than $ \mathcal{H}$ but of the same order of magnitude.
Second, let us take a close look at the nominator of (\ref{eq:av Xi1}), taking the form $k' \cdot P(t) = \omega' \mathcal{H} (1-\cos \theta_e)$ where $\theta_e$ is the angle between $\textbf{k}'$ and $\textbf{v}$. Since $\omega' \mathcal{H}$ and the denominator are of the same order of magnitude, it is clear that $\Xi_1 \ll 1$ is obtained
only at $\theta_e \ll 1$,
when $k' \cdot P(t) \approx \frac{1}{2}\omega' \mathcal{H}  \theta_e^2$.

Baier and Katkov \cite{katkov1} argued that for ultrarelativistic particles
($\gamma = \mathcal{H}/ m\gg 1$) the main contribution to the emission originates from the part of the trajectory where $\theta_e < 1/\gamma$. Hence, on the radiation formation region
$\Xi_1 \ll 1$ and $\mathcal{H}$ may be regarded as constant, and 
the conditions of Eqs.~(\ref{eq:av Xi1}), (\ref{eq:av Xi2}) are fulfilled.
We argue, however, that even though the main contribution originates from this part of the trajectory, Eqs.~(\ref{eq:av Xi1}), (\ref{eq:av Xi2}) should hold
on the entire cycle. The reason is that in a periodic motion, the transition amplitude contains a series of sequential contributions which interfere with each other generating the harmonic structure. In order to reproduce the correct phase between these contributions, $\psi$ should be valid through the entire cycle and not only on the emission regions.
The condition corresponding to this requirement may be derived as follows.
The particle's trajectory lies within a cone with angle $\sim \xi/\gamma$. Since, as established above, the emission angle is $\sim 1/\gamma$, the wave vector of the emitted photon $\textbf{k}'$ is restricted by a cone whose angle is $(1+\xi) / \gamma$.
Accordingly, $\theta_e \lesssim (1+\xi)/ \gamma$ and one may deduce that $\Xi_1 \approx O([1+\xi]/\gamma)^2$. On the other hand, the relative deviation of the energy is $\Xi_2 \approx O(\Delta \mathcal{H}/\mathcal{H}_{av}) \approx O(\xi/\gamma)$. As a result, the required conditions Eqs.~(\ref{eq:av Xi1}), (\ref{eq:av Xi2}) are satisfied for $ \gamma \gg 1+\xi $, which 
is more restrictive as compared to the 
common SC condition ($\gamma \gg 1$).

Let us calculate the intensity emitted by an electron in a REF with vanishing initial momentum ($\textbf{p} = 0$) corresponding to a circular classical trajectory. This initial condition is chosen for the following reasons. First, the simple classical motion allows for analytical expressions.
%, as we show below. 
Second, in this case $\gamma=\xi \gg 1$, so that the original Baier-Katkov condition is fulfilled, but our new restrictive one discussed above is not. In this way, our argument could be put to a test.
Thirdly, a circular trajectory corresponds as well to an electron interacting with PW in the case of a certain initial momentum choice \cite{Supp}.
As a consequence, we may compare the emission predicted by the full quantum calculation for two different field configurations sharing the same classical trajectory.
\begin{figure*}
  \begin{center}     	
  \includegraphics[width=0.9\textwidth]{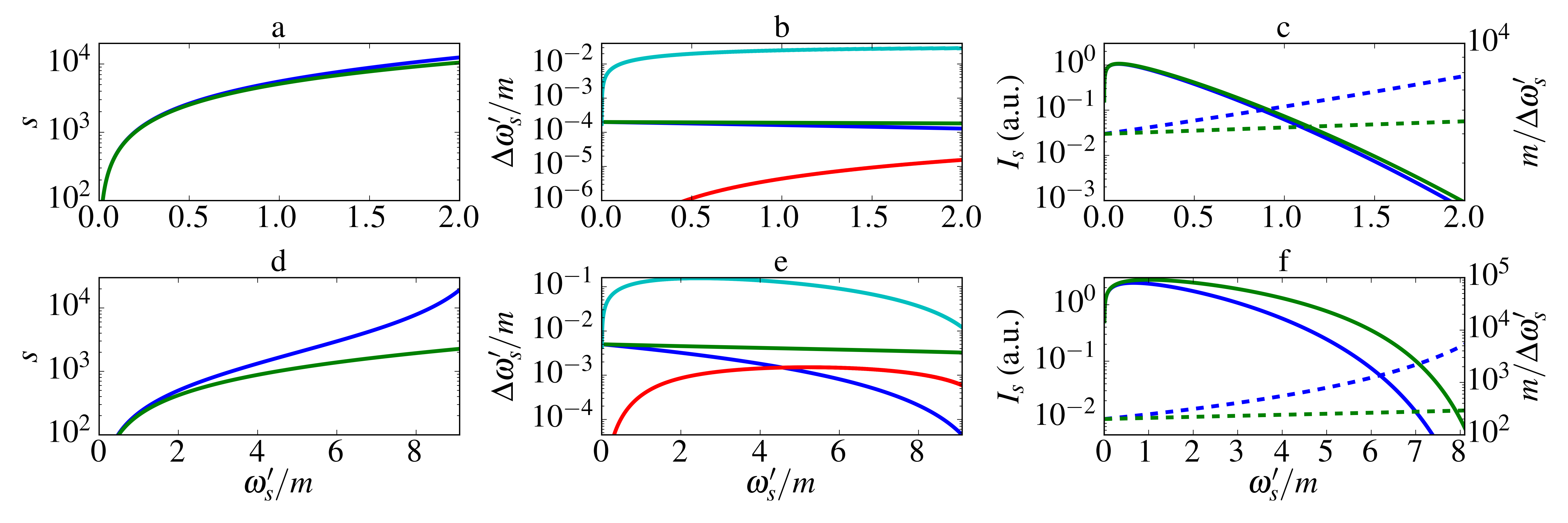}
      \caption{(a) The number of absorbed photons $s$ vs the emitted photon energy $\omega'_s$: (green) WKB and (blue) SC. The PW and SC curves coincide. (b) The width of the angle-integrated harmonics: (red) WKB, and (turquoise) PW (SC width is vanishing). The gap between neighboring harmonics: (green) WKB, and (blue) SC/PW.
      (c) The angle-integrated spectral intensity (solid) and the harmonic density (dashed): (green) WKB, (blue) SC/PW.
      The laser parameters are $\xi = 10, \omega = 100eV$, correseponding to $\chi = 0.02, I_L=10^{24}W/cm^2$. The electron initial momentum is $\textbf{p} = 0$. Subfigures (d,e,f) are similar to (a,b,c) but with $\omega = 2.55keV$, correseponding to $\chi = 0.5, I_L=10^{27}W/cm^2$.   }
      \label{fig}
  \end{center}
\end{figure*}
Due to the periodicity, one may decompose the phase into $\psi = \psi_p + \psi_{np} \omega t$, where $\psi_p,\psi_{np} \equiv \frac{1}{T} \int_0^{T} \psi(t')dt'$ are the periodic and non-periodic parts correspondingly and $T$ is the period of the motion.
Consequently, Eq.~(\ref{eq:av trans_amp12}) may be written as a series of delta function $\mathcal{T}_{fi}
%{-\infty}^{\infty}  
= 2 \pi  \sum_s\mathcal{M}_s  \delta \left( \Omega_s    \right)$, where $\Omega_s \equiv  \omega (s - \psi_{np})  = 0$ determines the harmonic energy $\omega_s'$. The matrix element corresponding to a $s$-photon process is given by
\begin{equation}
\mathcal{M}_s \equiv \frac{e}{T}\int_0^{T} d t \left[ \frac{\epsilon \cdot P}{\sqrt{2\omega' \mathcal{H} \mathcal{H}'}} \right] e^{i \psi_p - s \omega t},
\label{eq:av Ms}
\end{equation}
where $p' \equiv |\textbf{p}'|$, and $\theta, \varphi$ are the axial and azimuthal of the emitted photon. Comparing (\ref{eq:av trans_amp12}) and (\ref{eq:av Ms}), one may observe that $\mathcal{M}_s = \frac{1}{T} \mathcal{T}_{0T}(\omega'_s,\cos \theta)$.
Integrating over the outgoing electron momentum and the scattered photon energy and azimuthal angle, one arrives at the emission intensity
\begin{equation}
\frac{dI_s}{d(\cos \theta)} =  \frac{\omega_s'^3}{2 \pi }
\left| \frac{d \Omega_s}{d \omega'} \right|^{-1}_{\omega' = \omega_s'}
\sum_{\epsilon}
\bigl| \mathcal{M}_s \bigr| ^2 .
\label{eq:av rate2}
\end{equation}
Averaging the phase $\psi$ over a cycle, $\psi_{np}$ is obtained, allowing us to write down $\Omega_s$ explicitly
\begin{equation}
\Omega_{s} =  s \omega+ \mathcal{H} - \omega' - \mathcal{H}'_{av} (\omega')
\label{eq:av Om_wkb}
\end{equation}
where $\mathcal{H} = m \sqrt{1+\xi^2}$, and $\mathcal{H}'_{av} \equiv \frac{1}{\tau}  \int_0^{\tau}{dt} \mathcal{H'}= \frac{2}{\pi }\sqrt{G} E_2(\mu)$, see \cite{mocken}.
$E_2(\mu)$ is the complete elliptical integral of the second kind, $\mu \equiv 4m \xi \omega' \sin \theta/G$,
and $G \equiv \omega'^2 + m^2 \left( 1 + \xi^2 \right) + 2 \omega' m \xi \sin \theta$.
Accordingly, the derivative of $\Omega_s$, appearing in (\ref{eq:av Om_wkb}), may be evaluated analytically.
In order to find $\omega'_s$, Eq.~(\ref{eq:av Om_wkb}) is solved numerically.

The formalism outlined above
applies for the SC case as well. The difference lies in the different phase $\psi \rightarrow \widetilde{\psi}$ and therefore $\Omega_s \rightarrow \widetilde{\Omega}_s$.
Since the classical trajectory is 
given by $\textbf{x}(t)= \frac{\xi}{ \omega \sqrt{1+\xi^2}} \left[  \sin (\omega t) \hat{x} - \cos (\omega t) \hat{y} \right]$, $\widetilde{\psi}$ takes an analytical form. Averaging it to obtain the non-periodic part, one may find
\begin{equation}
\widetilde{\Omega}_s = s \omega - \frac{\omega' \mathcal{H}}{\mathcal{H} - \omega'}
\label{eq:av Om_SC}
\end{equation}
As a result, the relation between the number of absorbed photons and the emitted photon energy reads $\omega'_s = 
s \omega \mathcal{H}/(s \omega + \mathcal{H})$.
A similar expression is obtained for the PW case \cite{Supp}.

From the derivation of the emission intensity above we can come to several conclusions. 
We may consider the low and high energy limits.
One may see that for $\omega' /\mathcal{H} \ll 1$, we have $\mathcal{H} \approx \mathcal{H}'_{av}$ so that $\Omega_s \approx \widetilde{\Omega}_s \approx s \omega - \omega'$, i.e., in the classical limit the harmonics are simply multiples of the REF frequency.
On the other hand, the high energy spectrum exhibits qualitatively different behavior.
The quantum model predicts a cutoff for the number of absorbed photons, whereas according to the SC approach $s$ tends, in principle, to infinity.
A crude estimation to this cutoff may be obtained as follows. The high energy tail of the spectrum corresponds to $\omega' \rightarrow \mathcal{H}$. Consequently, $\Omega_s$ approximately reduces to $s_{c} \omega \approx \mathcal{H}'_{av}
$.
In this limit $\mu \rightarrow 1$, so that $\mathcal{H}'_{av} = 4m \xi E_2(1)/ \pi$. Since $E_2(1) = 1$, the cutoff may be estimated as
\begin{equation}
s_{c}  \approx \frac{4m \xi }{ \pi \omega}.
\label{eq:av s_c}
\end{equation}
The dimensionless parameter $s_c$ determines the maximal number of absorbed photons for the REF configuration. 
For the SC and PW cases, on the other hand, the emission probability indeed decays for $\omega' \rightarrow \mathcal{H}$, but such a restriction on $s$ does not exist, giving rise to the emergence of a semi-continuum.

Further, a close look reveals that a relation between the matrix elements corresponding to the two approaches may be established. As explained above, the difference between $\mathcal{T}, \widetilde{\mathcal{T}}$ originates from the interference between different cycles.
The matrix element integration, however, is limited to a single cycle and a single saddle point contributing to the emission. Accordingly, the amplitudes of the two methods are sampling of the same function, where the difference arises from the different sampling frequency, namely the harmonics $\omega'_s,\widetilde{\omega}'_s$ respectively \cite{Supp}:
\begin{equation}
\mathcal{T}_{0T}(\omega',\cos \theta)\approx
\widetilde{\mathcal{T}}_{0T}(\omega',\cos \theta)
\label{eq:av tau_eq}
\end{equation}

Finally, one may deduce from Eq.~(\ref{eq:av Om_SC}) that the SC harmonics have no angular dependence,
since $\widetilde{\Omega}_s$ bears no angular dependence. Consequently, after integration over the angular distribution, the width of the harmonics is still vanishing. For the WKB method, however, $\mathcal{H}'_{av}$ is angle dependent. In the case examined numerically below, however, it is explicitly shown that the width after angle integration is much smaller as compared to the spacing between neighboring harmonics, so that they may be regarded as discrete as well.

In the following, the emission properties are calculated numerically via the quantum WKB and SC methods for the REF configuration, as well as compared with the well-known PW result.
The non-linear parameter is $\xi = 10$ and the electron initial momentum is $\textbf{p} = 0$.
Two cases were considered, corresponding to different values of the quantum parameter $\chi = \xi^2 (\omega/m)$. The higher the $\chi$, the larger the discrepancy between WKB and SC.
The case of $\chi=0.5$, presented in Fig.~\ref{fig}(e,d,f),  illustrates the expected effect in the spectral distribution.
It may be realized with $\omega = 2.55keV$,
corresponding to a XFEL laser with an intensity of about $I_L \approx  10^{27}W/cm^2$.
Smaller but still visible effect can be obtained, as shown in Fig.~\ref{fig}(a,b,c), even by using less demanding conditions
$\omega=100eV, \chi = 0.02,I_L \approx  10^{24}W/cm^2$.
These intensities lay above present day achievable intensity \cite{LCLS,XFEL,nature_compton}.
Nevertheless, improvements of the focusing technique to approach the diffraction limit may allow for such intensities in the future \cite{ringwald}. 

Fig.~\ref{fig}(a,d) show the relation between the number of absorbed laser photons $s$ and the emitted photon energy $\omega'_s$ for the lower and higher intensity respectively. The PW and SC predictions are identical, as also shown analytically in \cite{Supp}.
For small values of the emitted photon energy, the SC and WKB curves coincide, as explained above. 
The deviation occurs near the high energy tail of the spectrum. For the lower intensity shown in Fig.~\ref{fig}(a), it amounts to about 20 percents for harmonics still having a significant intensity (see Fig.~\ref{fig}(c)). 
For the high intensity case depicted in Fig.~\ref{fig}(d), the discrepancy is of orders of magnitude and the different asymptotic behavior for $\omega' \rightarrow \mathcal{H}$ is manifest. One may see that according to the SC model, $s$ asymptotically increases, while the WKB model predicts a finite cutoff.
The cutoff value is $s_c \approx 2550$, in agreement with the theoretical estimation of Eq.~(\ref{eq:av s_c}).
The physical meaning is that in a REF, higher amount of energy would be depleted from the external field as compared to the same emission in a PW. Furthermore, since the difference between the absorbed laser energy $s \omega$ and the emitted photon energy $\omega'$ is converted to the kinetic energy of the electron, it implies that for the PW the emission is accompanied by higher electron acceleration.

The gap between harmonics $\Delta \omega'_s$ for the various approaches, as well as the spectral width of the angle-integrated spectrum associated with the WKB and PW harmonics (defined such that 2/3 of the energy is contained within this width) are shown in Fig.~\ref{fig}(b,e). Note that the angle-integrated SC harmonics have no width, as shown analytically above.
First, one may observe that the gap changes only slightly for the WKB, but decreases significantly for the SC (40 and 1.4 times for the high and low intensity respectively).
Second, the WKB harmonics width is significantly lower than the gap, implying that they may be regarded as discrete. 
The PW width, however, is always much higher than the gap, so that the angle-integrated spectrum is continuous and the harmonics cannot be distinguished. Thus, the quantum mechanical calculation of emission corresponding to two distinct field configurations yields different results, even though the associated classical trajectories are similar. In other words, when the 
condition pointed by us above, $\gamma \gg 1+ \xi$, is violated, the classical trajectory does not solely determine the emission, but rather the dispersion relation of the external field photons plays a role as well. 

The angle-integrated spectral intensity of harmonics is presented in Fig.~\ref{fig}(c,f).
The PW and SC curves coincide, as in Fig.~\ref{fig}(a,d). One can see that the high energy WKB harmonics are stronger than the SC ones (40 and 1.4 times for the high and low intensity respectively).
This figure shows also the normalized harmonics density $m/\Delta \omega'_s$.
The latter indicates clearly that for the SC the harmonics density is much higher. In other words,
the WKB exhibits distant and intense harmonics, while the SC predicts weak and spectrally dense ones. Averaging the discrete harmonics (i.e. dividing their power by $\Delta \omega'_s$), one obtains exactly the same spectrum for all models, coinciding with the CCF prediction as well.
The equivalent total energy may be intuitively explained as follows.
From Eq.~(\ref{eq:av tau_eq}) it follows that for a single cycle pulse the quantum and SC method should yield exactly the same spectrum. The transition from a single cycle to a periodic pulse gives rise to the concentration of the emitted energy in discrete harmonics (which are different for the two models as seen above) but does not change the total energy, see \cite{Supp}.

Concluding, we show that the well-known Baier-Katkov SC formula for radiation of relativistic electrons is valid only when an additional  condition of $\gamma \gg 1+\xi$ is fulfilled.
Considering a particular case $\gamma = \xi \gg 1$ when the additional condition is violated, discrepancy between the SC and quantum spectral harmonic structure is demonstrated, whereas the harmonic-averaged spectrum remains equivalent. In particular, the quantum calculation predicts a maximal value for the possible number of absorbed photons, while according to the SC model it is unlimited. Furthermore, comparing the quantum calculations corresponding to the REF and PW, it was observed that two various field configurations with the same classical trajectory may result in different emission characteristics due to different dispersion relations of the strong field photons.

The authors are greatful to Antonino Di Piazza for helpful discussions.
ER acknowledges support from the Alexander von Humboldt Foundation.

%\clearpage
%\newpage\newpage
\clearpage


\section*{Supplementary material}

# Semi-classical limitations for photon emission in strong external fields: Supplemental Material

## TRANSITION AMPLITUDE

The wave function equation of motion for a scalar particle in a strong electromagnetic (EM) field reads [1]

$$\left[-\hbar^2 \partial^2 - 2ie\hbar(A \cdot \partial) + e^2 A^2 - m^2\right] \Phi = 0 \quad (1)$$

where $\hbar$ is explicitly written. The vector potential corresponding to a rotating electric field, as defined in the main text, reads $A^\mu = a_1^\mu \cos(\omega t) + a_2^\mu \sin(\omega t)$, where $a_1^\mu = a(0, 1, 0, 0)$, $a_2^\mu = a(0, 0, 1, 0)$ are the polarization vectors. We employ the leading order WKB method, i.e. we seek for a solution of the form

$$\Phi = \exp\left[\left(\frac{S_0}{\hbar} + S_1\right)\right]. \quad (2)$$

Substituting (2) into (1), neglecting the second order terms in $\hbar$ and equating the coefficients of each $\hbar$ power to zero, one finds

$$S_0 = i\left(\int_{-\infty}^t dt' \mathcal{H}(t') - p \cdot x\right), \quad S_1 = -\frac{1}{2}\ln\mathcal{H}(t), \quad (3)$$

where $p_\mu$ is the initial momentum and $S_0$ is the classical action, as expected. From now on the $\hbar$ will be omitted. Accordingly, the approximated wave function reads

$$\Phi = \frac{1}{\sqrt{2\mathcal{H}(t)}} \exp\left[i\left(\int_{-\infty}^t dt' \mathcal{H}(t') - p \cdot x\right)\right]. \quad (4)$$

In principle, a factor of $1/\sqrt{V}$ should be introduced, where V is the normalization volume. However, since it cancels out in the rate calculation detailed below, it will be omitted from now on. Notice that for a general EM field configuration, the classical action and trajectory cannot be found analytically and therefore the WKB is not beneficial. In our case, however, the classical motion admits an analytical solution.

The transition amplitude for the leading order non-linear photon emission, depicted as a Feynmann diagram in Fig.1, is given [1] by

$$iT = e\epsilon^\mu \int dx^4 \frac{e^{-ik' \cdot x}}{\sqrt{2\omega'}}$$
$$\left[\Phi'^* (\partial_\mu + eA_\mu) \Phi - \Phi' (\partial_\mu - eA_\mu) \Phi^*\right], \quad (5)$$

where the $'*'$ symbol stands for a complex conjugate. Substituting the wavefunction (2) leads to

$$iT = (2\pi)^3 e\epsilon^\mu \int_{-\infty}^\infty dt$$
$$\left[\frac{P_\mu(t) + P'_\mu(t)}{\sqrt{8\omega' \mathcal{H}(t)\mathcal{H}'(t)}}\right] e^{i\psi} \delta^3(\mathbf{p} - \mathbf{p}' - \mathbf{k}'), \quad (6)$$

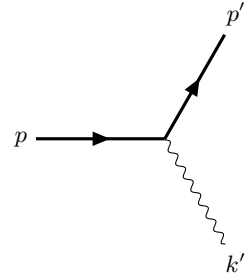

FIG. 1. The Feynmann diagram representing the emission process. The bold lines represent the "dressed" electron in the Furry picture [2].

where $P_\mu(t)$ is the kinetic momentum and the argument reads

$$\psi = \omega' t + \int_\infty^t dt' \mathcal{H}'(t') - \int_{-\infty}^t dt' \mathcal{H}(t'). \quad (7)$$

It follows from the delta function in (6) that the initial momenta of the in-coming and out-coming particles are related through $\mathbf{p}' = \mathbf{p} - \mathbf{k}'$. As a result, from $\mathbf{P}(t) = \mathbf{p} - e\mathbf{A}(t)$ it may be deduced that $\mathbf{P}' = \mathbf{P} - \mathbf{k}'$. Since the temporal component of $\epsilon_\mu$ vanishes, we have

$$\epsilon \cdot [P(t) + P'(t)] = -\epsilon \cdot \left[2\mathbf{P}(t) - \mathbf{k}'\right] = 2\epsilon \cdot P(t), \quad (8)$$

where the orthogonality relation $\epsilon \cdot k' = 0$ was used. Finally, one obtains

$$iT = (2\pi)^3 \mathcal{T}_{fi} \delta^3(\mathbf{p}_c) \quad (9)$$

where $\mathbf{p}_c \equiv \mathbf{p} - \mathbf{p}' - \mathbf{k}'$ and

$$\mathcal{T}_{fi} \equiv e\int_{t_i}^{t_f} dt' \frac{[\epsilon \cdot P(t')]}{\sqrt{2\omega' \mathcal{H}(t')\mathcal{H}'(t')}} e^{i\psi}. \quad (10)$$

## RATE CALCULATION

The transition amplitude is related to the emission probability $\mathcal{P}$ through [3]

$$\mathcal{P} = \int \frac{Vd^3p'}{(2\pi)^3} \frac{Vd^3k'}{(2\pi)^3} |iT|^2. \quad (11)$$

The squared transition amplitude (9) takes the form

$$|iT|^2 = (2\pi)^3 \delta^3(\mathbf{p} - \mathbf{p}' - \mathbf{k}') V |\mathcal{T}_{fi}|^2. \quad (12)$$

Plugging (12) into (11) and taking advantage of the spatial $\delta$ function one arrives at

$$\mathcal{P} = \frac{1}{(2\pi)^3} \int d\varphi d\cos(\theta) d\omega' \omega'^2 |\mathcal{T}_{fi}|^2, \quad (13)$$

where $\theta$ and $\varphi$ are the axial and azimuthal angles corresponding to the out-coming photon. The volume $V$ is cancelled with the normaliztion factor of the wavefunctions included in $\mathcal{T}_{fi}$. Squaring $\mathcal{T}_{fi}$ yields

$$|\mathcal{T}_{fi}|^2 = 2\pi \sum_s |\mathcal{M}_s|^2 \delta(\Omega_s)\tau, \qquad (14)$$

where the matrix element and the $\delta$ function argument, respectively, are defined as

$$\mathcal{M}_s = \frac{1}{T}\int_0^T dt \left[\frac{\epsilon \cdot P(t)}{\sqrt{2\mathcal{H}\mathcal{H}'\omega_s'}}\right] e^{i\psi_p - s\omega t}, \qquad (15)$$

and

$$\Omega_s = \mathcal{H} + s\omega - \omega' - \mathcal{H}'_{av}(\omega'). \qquad (16)$$

$\tau$ is the interaction time which goes to infinity since we assume that the pulse is very long. Introducing the probability per time unit, $W \equiv \lim_{\tau \to \infty} \frac{1}{\tau}\mathcal{P}$, we obtain

$$\frac{dW}{d(\cos\theta)} = \frac{1}{(2\pi)^2}\int d\omega'\omega'^2 \sum_s |\mathcal{M}_s|^2 \delta(\Omega_s). \qquad (17)$$

Notice that owing to the azimuthal symmetry, the transition amplitude does not depend on $\varphi$, so it can be integrated out. Employing the relation between the intensity and the probability per unit time, namely $dI_s = \omega_s' dW_s$, leads to the final result

$$\frac{dI}{d(\cos\theta)} = \frac{1}{2\pi}\omega'^3 \sum_s |\mathcal{M}_s|^2 \left|\frac{d\Omega_s}{d\omega'}\right|^{-1}_{\omega'=\omega_s'}. \qquad (18)$$

### MATRIX ELEMENT CALCULATION

#### SC approach

The semi-classical matrix element is defined in the main text by

$$\widetilde{\mathcal{M}}_s = \frac{1}{T}\int_0^T dt \left[\frac{\epsilon \cdot P(t)}{\sqrt{2\mathcal{H}(\mathcal{H} - \widetilde{\omega}_s')\widetilde{\omega}_s'}}\right] e^{i\widetilde{\psi}_p - s\omega t}, \qquad (19)$$

where the periodic part of the SC argument is

$$\widetilde{\psi}_p = \frac{\mathcal{H}}{\mathcal{H} - \widetilde{\omega}_s'}\mathbf{k}' \cdot \mathbf{x}(t). \qquad (20)$$

Without loss of generality, we consider a photon emitted in the $x-z$ plane, i.e. with $\varphi = 0$. Hence, $\mathbf{k}' \cdot \mathbf{x}(t) = \widetilde{\omega}_s' x(t)\sin\theta$ where $x(t)$ is the $x$ component of the trajectory. It is given by $x(t) = \frac{\xi}{\omega\sqrt{1+\xi^2}}\sin(\omega t)$. In addition, as shown in the main text, the relation between the harmonic index $s$ and the emitted photon energy reads

$$\widetilde{\omega}_s' = \frac{s\omega\mathcal{H}}{s\omega + \mathcal{H}}. \qquad (21)$$

Substituting Eq. (21) as well as $x(t)$ into (20) we obtain

$$\widetilde{\psi}_p = \frac{s\xi}{\sqrt{1+\xi^2}}\sin\theta\sin(\omega t) \equiv z\sin(\omega t), \qquad (22)$$

where we have defined

$$z \equiv \frac{s\xi\sin\theta}{\sqrt{1+\xi^2}}. \qquad (23)$$

Since $P_\mu = (\mathcal{H}, m\xi\cos(\omega t), m\xi\sin(\omega t), 0)$, the matrix element reads

$$\widetilde{\mathcal{M}}_s = \epsilon_\mu \widetilde{\mathcal{M}}_s^\mu \qquad (24)$$

where

$$\widetilde{\mathcal{M}}_s^\mu \equiv \frac{m}{\sqrt{2\mathcal{H}(\mathcal{H}-\widetilde{\omega}_s')\widetilde{\omega}_s'}}(B\sqrt{1+\xi^2}, \xi B_1, \xi B_2, 0). \qquad (25)$$

The coefficients $B, B_1, B_2$ are defined as

$$B \equiv \frac{1}{2\pi}\int_0^{2\pi} d\phi\, e^{i(z\sin\phi - s\phi)}, \qquad (26)$$

$$B_1 \equiv \frac{1}{2\pi}\int_0^{2\pi} d\phi\, \cos\phi\, e^{i(z\sin\phi - s\phi)}, \qquad (27)$$

and

$$B_2 \equiv \frac{1}{2\pi}\int_0^{2\pi} d\phi\, \sin\phi\, e^{i(z\sin\phi - s\phi)}, \qquad (28)$$

where $\phi = \omega t$. Recalling the integral definition of the Bessel function $J_s$ one may write

$$B = J_s(z), \quad B_1 = \frac{s}{z}J_s(z), \quad B_2 = -iJ_s'(z). \qquad (29)$$

Due to the Ward indentity [3], the sum over the photon polarizations is given by

$$\sum_\epsilon |\widetilde{\mathcal{M}}_s|^2 = -\widetilde{\mathcal{M}}_s^\mu \widetilde{\mathcal{M}}_{s\mu}. \qquad (30)$$

Employing (24-29) we obtain

$$\sum_\epsilon |\widetilde{\mathcal{M}}_s|^2 = K\left(\xi^2\left[\left(\frac{s^2}{z^2}-1\right) + J_s'^2(z)\right] - J_s^2(z)\right), \qquad (31)$$

where

$$K \equiv \frac{m^2}{2\mathcal{H}(\mathcal{H}-\widetilde{\omega}_s')\widetilde{\omega}_s'}. \qquad (32)$$

The final result is very similar to the matrix element corresponding to a circularly polarized plane wave, as obtained by Ritus and Nikishov [4] (see also the plane

wave section below). The main difference lies within the definition of $z$.

Since we are interested in the $\xi \gg 1$ domain, the integrals appearing in (26-28) are rapidly oscillating and therefore may be approximated using the saddle point technique, yielding [4]

$$J_s(z) \approx \left(\frac{2}{s}\right)^{1/3} Ai(y), \quad J'_s(z) \approx \left(\frac{2}{s}\right)^{2/3} Ai'(y), \quad (33)$$

where $Ai$ designates the Airy function and its argument $y$ is related to $z, s$ through

$$y \equiv \left(\frac{s}{2}\right)^{2/3}\left(1 - \frac{z^2}{s^2}\right). \quad (34)$$

As a result, the sum over polarizations of the squared matrix element reads

$$\sum_\epsilon |\widetilde{\mathcal{M}}_s|^2 \approx$$
$$K\left(-Ai^2(y) + \xi^2 \left(\frac{2}{s}\right)^{4/3}\left[yAi^2(y) + Ai'^2(y)\right]\right). \quad (35)$$

### WKB quantum approach

In the following, the relation between the quantum and SC matrix elements is established. The WKB matrix element is defined in the main text as follows

$$\mathcal{M}_s = \frac{1}{T}\int_0^T dt \left[\frac{\epsilon \cdot P(t)}{\sqrt{2\mathcal{H}\mathcal{H}'\omega'_s}}\right] e^{i\psi_p - s\omega t}. \quad (36)$$

Notice that as opposed to $\mathcal{H} - \widetilde{\omega}'_s$ appearing in the denominator of (19), (36) contains instead the time-dependent quantity $\mathcal{H}'$. The periodic part of the phase takes the form

$$\psi_p = \mathcal{H}'_{av}t - \int_{-\infty}^t dt' \mathcal{H}'. \quad (37)$$

For the case considered in the main text, i.e. $\gamma \approx \xi \gg 1$, the argument $\psi$ is rapidly oscillating and the contribution to the integral comes from the region where $\theta_e < 1/\gamma$, as discussed in the main text. Therefore, one may use the Baier-Katkov approximation for the phase

$$\omega'_s t + \int_{-\infty}^t dt' \mathcal{H}' - \mathcal{H} \approx \int_{-\infty}^t dt' \frac{\mathcal{H}}{\mathcal{H} - \omega'_s}\left(k'_s \cdot v\right), \quad (38)$$

where $k'_s = \omega'_s(1, \sin\theta, 0, \cos\theta)$ is the wave vector of the emitted photon corresponding to the $s$ harmonic. Plugging $\int_\infty^t dt' \mathcal{H}'$ from (38) into (37) and using $s\omega + \mathcal{H} = \omega'_s + \mathcal{H}'_{av}$ one obtains

$$\psi_p = \mathcal{H}'_{av}t - \mathcal{H}t - \int_{-\infty}^t dt' \frac{\mathcal{H}}{\mathcal{H} - \omega'_s}\left(k'_s \cdot v\right) + \omega'_s t =$$
$$s\omega - \int_{-\infty}^t dt' \frac{\mathcal{H}}{\mathcal{H} - \omega'_s}\left(k'_s \cdot v\right). \quad (39)$$

Hence, in the vicinity of the saddle point we have

$$\psi_p - s\omega t \approx -\int_{-\infty}^t dt' \frac{\mathcal{H}}{\mathcal{H} - \omega'_s}\left(k'_s \cdot v\right). \quad (40)$$

Writing explicitly the right wing we have

$$\psi_p - s\omega t \approx \frac{\mathcal{H}}{\mathcal{H} - \omega'_s}\mathbf{k}'_s \cdot \mathbf{x}_p - s_{eff}\omega t, \quad (41)$$

where $s_{eff}$ as defined by

$$s_{eff}\omega \equiv \frac{\omega'_s \mathcal{H}}{\mathcal{H} - \omega'_s} \quad (42)$$

is an effective (and not necessarily integer) index for which the emitted energy of the SC model coincides with the WKB harmonic $\omega'_s$. One may explicitly verify that even though the functions on the right and left wings of (41) are not identical, their Taylor expansion up to third order in the vicinity of the saddle point is identical. Additionally, it should be noted that the prefactor $1/\sqrt{\mathcal{H}'(t)}$ is time dependent, as opposed to the SC case. However, since a detailed examination shows that its derivative with respect to $t$ vanishes at the saddle point $t = 0$, it may be considered as constant. Hence, as both the prefactor and the phase of the WKB and SC are identical near the saddle point, one may write

$$\mathcal{M}_s \approx \widetilde{\mathcal{M}}_{s_{eff}}. \quad (43)$$

Namely, the WKB matrix element coincides with the SC one corresponding to an effective harmonic index $s_{eff}$, so that the emitted energy $\omega'$ would be the same. Finally, from the above derivation and due to the relation

$$\mathcal{M}_s = \frac{1}{T}\mathcal{T}_{0T}(\omega'_s, \cos\theta), \quad (44)$$

presented in the main text, it follows that

$$\mathcal{T}_{0T}(\omega', \cos\theta) \approx \widetilde{\mathcal{T}}_{0T}(\omega', \cos\theta). \quad (45)$$

The angular dependence was explicitly written for the sake of clarity.

### PLANE WAVE AND CONSTANT CROSSED FIELD FORMULAS

First, let us obtain the initial momentum for a particle counterpropagating a plane wave which results in a circular motion, similar to the REF case. We recall that in this case the time-dependent momentum is given by [4]

$$P_\mu = p_\mu - eA_\mu + \frac{e^2 a^2}{2(k^l \cdot p)}k^l_\mu, \quad (46)$$

where $k^l = (\omega, 0, 0, \omega)$. As a result, for initial momentum $(p_0, 0, 0, p_z)$ satisfying $p_z = e^2 a^2/[2(p_0 - p_z)]$, the

$z$ component of the momentum vanishes leading to the same classical trajectory (circle) and quantum parameter $\chi$ as in the REF problem we consider. As a consequence, we may compare the emission predicted by full quantum calculations for two different EM configurations sharing the same classical trajectory.

The intensity of the radiation emitted by a scalar particle interacting with a circularly polarized plane wave is given by [4]

$$\frac{dI}{du} = \frac{e^2 m^2}{2\pi} \frac{u}{(1+u)^3} \sum_s \left[ -J_s^2(z) + \xi^2 \left[ \left(\frac{s^2}{z^2} - 1\right) J_s^2(z) + J_s'^2(z) \right] \right], \quad (47)$$

where the relation between $u$ and the emitted photon energy for ultra-relativistic particles reads

$$u = \frac{\omega'}{\mathcal{H} - \omega'}, \quad (48)$$

so that $\omega' = u\mathcal{H}/(1+u)$. The Bessel function argument is given by

$$z \equiv \frac{\xi^2 \sqrt{1+\xi^2}}{\chi} \sqrt{u(u_s - u)}. \quad (49)$$

The quantum parameter $\chi$ is defined in the main text. Each harmonic is non-vanishing on the interval $0 < u < u_s$ where

$$u_s \equiv \frac{2s\chi}{\xi(1+\xi^2)}. \quad (50)$$

The harmonic spectral location is defined according to the peak emission, corresponding to $u = u_s/2$. Using (48) one may write it in terms of the emitted photon energy

$$\omega'_s = \frac{u_s}{2(1 + u_s/2)} \mathcal{H}. \quad (51)$$

Employing (50) as well as $\chi = \frac{\xi \mathcal{H}\omega}{m^2}$ and $\mathcal{H} = m\sqrt{1+\xi^2}$ one obtains

$$\omega'_s = \frac{s\omega \mathcal{H}}{s\omega + \mathcal{H}}. \quad (52)$$

Notice that this expression is identical to the one obtained for the SC model in the main text.

The Constant-Crossed-Field expression arises from (47) by using the saddle point analysis (similar to the derivation in the SC matrix element section) and replacing the discrete harmonics with a continuous variable, namely

$$\frac{dI}{du} = \frac{e^2 m^2}{2\pi} \frac{u}{(1+u)^3} \left(\frac{u}{2\chi}\right)^{1/3} \int_{-\infty}^{\infty} d\tau \left[ -Ai^2(y) + \left(\frac{2\chi}{u}\right)^{2/3} \left[ y Ai^2(y) + Ai'^2(y) \right] \right], \quad (53)$$

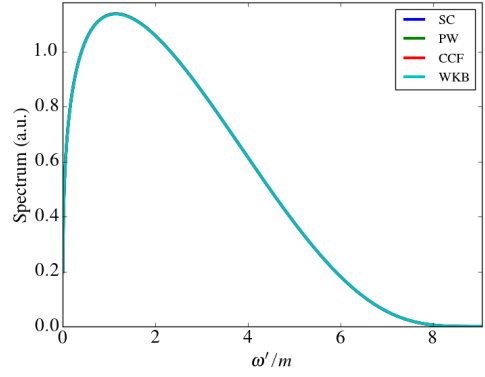

FIG. 2. The harmonics-averaged spectrum: (blue) SC, (green) PW and (turquise) WKB as compared to the CCF expression. All curves coincide to a very good approximation.

where

$$y \equiv \left(\frac{u}{2\chi}\right)^{2/3} \left[1 + \tau^2\right]. \quad (54)$$

### AVERAGE SPECTRUM EQUIVALENCE

In the SC matrix element section the relation (45) between $\mathcal{T}_{0T}, \widetilde{\mathcal{T}}_{0T}$ was established. It follows that the transition amplitude is the same for the WKB and SC approaches given that the EM field contains a single cycle only. In the following, we argue that owing to (45), one may expect that the harmonic-averaged spectrum predicted by the two models should coincide for a periodical field. We notice that $\mathcal{T}_{fi}$ may be cast in the form

$$\mathcal{T}_{fi} = \int_{-\infty}^{\infty} dt f(t, \omega') e^{i\omega' t}, \quad (55)$$

where

$$f(t, \omega') \equiv \frac{[\epsilon \cdot P(t)]}{\sqrt{2\omega' \mathcal{H}(t) \mathcal{H}'(t)}} \exp\left[\left(\int_{\infty}^{t} dt' \mathcal{H}'(t') - \int_{-\infty}^{t} dt' \mathcal{H}(t')\right)\right]. \quad (56)$$

It resembles a Fourier transform except for the fact that $f(t, \omega')$ depends on $\omega'$ as well as on $t$. Furthermore, according to the derivation above (see rate calculation section), the intensity is proportional to

$$\frac{dI}{d\cos\theta} \propto \int_{-\infty}^{\infty} \omega'^3 |\mathcal{T}_{fi}|^2 d\omega', \quad (57)$$

which takes the same form as the energy integration over a Fourier spectrum, except for the $\omega'^3$ factor. In order

to take advantage of this analogy we recall one of the Fourier transform features. Let us consider a function $q(l)$ which is non-vanishing only in the interval $0 < l < L$. Its Fourier transform reads

$$Q(v) = \int_0^L q(l) e^{-ivl} dl. \tag{58}$$

Now we use $q(l)$ to define a periodic function

$$q_p(l) = q(l - nL), \qquad n \equiv \left\lfloor \frac{l}{L} \right\rfloor, \tag{59}$$

where $\lfloor \rfloor$ designates the flooring function. The Fourier transform of the new function is therefore

$$Q_p(v) = \sum_j \delta(v - v_j) Q(v_j), \qquad v_j \equiv \frac{2\pi j}{L}. \tag{60}$$

Namely, the transition from a single to periodic function induced "roaming" of the spectral energy to discrete frequencies (harmonics) without changing the average enrgy. In order to observe it explicitly, let us consider the enrgy contained in the interval $v_j < v < v_{j+1}$

$$\int_{v_j}^{v_{j+1}} dv |Q_p(v)|^2 =$$
$$\frac{1}{2} \left(Q(v_j) + Q(v_{j+1})\right) [v_{j+1} - v_j]. \tag{61}$$

One may see that (61) equals approximately the integral over a single cycle spectrum $\int_{v_j}^{v_{j+1}} dv |Q(v)|^2$, given that the envelope function $Q(v)$ changes only slightly in the interval under consideration.

The same reasoning applies in our case, given that the functions $\omega'^3$ and $f(t, \omega')$ are slowly varying envelope functions which stay almost constant on the frequency scale of the above mentioned "energy roaming", i.e. $\omega' \to \omega' + \Delta\omega'_s$. Since $\Delta\omega'_s \leq \omega$, as demonstrated in Fig. 1(b,e) in the main text, and for $\xi \gg 1$ corresponding to $\omega' \gg \omega$, this condition is satisfied and hence one should expect the harmonic-averaged spectrum to be equivalent to the single cycle one. The same argument applies for the SC case, so that its harmonic-averaged spectrum recovers the single cycle one. Finally, since the WKB and SC single cycle spectrum is equivalent, as established above, one may deduce that the harmonic-averaged spectrum of the WKB and SC approaches coincide as well. Fig. 3 shows the harmonic-averaged spectrum of the WKB and SC models, as well as the PW and CCF spectrum. One may see that all expressions coincide to a very good approximation.



\end{document}